\documentclass{aa}
\usepackage{graphicx}
\usepackage{txfonts}
\usepackage{natbib}
\usepackage[figuresright]{rotating}
\usepackage{color}
\usepackage{balance}
\idline{15}{66}

\begin{document}

\def\teff{$T\rm_{eff}$ }
\def\kms {\,$\mathrm{km\, s^{-1}}$ }
\def\kmss {\,$\mathrm{km\, s^{-1}}$}
\def\ms {$\mathrm{m\, s^{-1}}$ }

\newcommand{\Teff}{\ensuremath{T_\mathrm{eff}}}
\newcommand{\g}{\ensuremath{g}}
\newcommand{\gf}{\ensuremath{gf}}
\newcommand{\loggf}{\ensuremath{\log\gf}}
\newcommand{\glog}{\ensuremath{\log\g}}
\newcommand{\pun}[1]{\,#1}
\newcommand{\cobold}{\ensuremath{\mathrm{CO}^5\mathrm{BOLD}}}
\newcommand{\linfor}{Linfor3D}
\newcommand{\xx}{\ensuremath{\mathrm{1D}_{\mathrm{LHD}}}}
\newcommand{\punms}{\mbox{\rm\,m\,s$^{-1}$}}
\newcommand{\punkms}{\mbox{\rm\,km\,s$^{-1}$}}
\newcommand{\abuhe}{\mbox{Y}}
\newcommand{\grav}{\ensuremath{g}}
\newcommand{\mlp}{\ensuremath{\alpha_{\mathrm{MLT}}}}
\newcommand{\mlpcm}{\ensuremath{\alpha_{\mathrm{CMT}}}}
\newcommand{\moh}{\ensuremath{[\mathrm{M/H}]}}
\newcommand{\senv}{\ensuremath{\mathrm{s}_{\mathrm{env}}}}
\newcommand{\shelio}{\ensuremath{\mathrm{s}_{\mathrm{helio}}}}
\newcommand{\smin}{\ensuremath{\mathrm{s}_{\mathrm{min}}}}
\newcommand{\spun}{\ensuremath{\mathrm{s}_0}}
\newcommand{\sstar}{\ensuremath{\mathrm{s}^\ast}}
\newcommand{\tauross}{\ensuremath{\tau_{\mathrm{ross}}}}
\newcommand{\ttaurelation}{\mbox{T$(\tau$)-relation}}
\newcommand{\Ysurf}{\ensuremath{\mathrm{Y}_{\mathrm{surf}}}}
\newcommand{\mD}{\ensuremath{\left\langle\mathrm{3D}\right\rangle}}

\newcommand{\draftflag}{false}

\newcommand{\beq}{\begin{equation}}
\newcommand{\eeq}{\end{equation}}
\newcommand{\pdx}[2]{\frac{\partial #1}{\partial #2}}
\newcommand{\pdf}[2]{\frac{\partial}{\partial #2}\left( #1 \right)}

\newcommand{\var}[1]{{\ensuremath{\sigma^2_{#1}}}}
\newcommand{\sig}[1]{{\ensuremath{\sigma_{#1}}}}
\newcommand{\cov}[2]{{\ensuremath{\mathrm{C}\left[#1,#2\right]}}}
\newcommand{\xtmean}[1]{\ensuremath{\left\langle #1\right\rangle}}

\newcommand{\eref}[1]{\mbox{(\ref{#1})}}

\newcommand{\Vact}{\ensuremath{\nabla}}
\newcommand{\Vad}{\ensuremath{\nabla_{\mathrm{ad}}}}
\newcommand{\Veddy}{\ensuremath{\nabla_{\mathrm{e}}}}
\newcommand{\Vrad}{\ensuremath{\nabla_{\mathrm{rad}}}}
\newcommand{\Vraddiff}{\ensuremath{\nabla_{\mathrm{rad,diff}}}}
\newcommand{\cp}{\ensuremath{c_{\mathrm{p}}}}
\newcommand{\taueddy}{\ensuremath{\tau_{\mathrm{e}}}}
\newcommand{\vconv}{\ensuremath{v_{\mathrm{c}}}}
\newcommand{\Fconv}{\ensuremath{F_{\mathrm{c}}}}
\newcommand{\lmix}{\ensuremath{\Lambda}}
\newcommand{\Hp}{\ensuremath{H_{\mathrm{P}}}}
\newcommand{\Hptop}{\ensuremath{H_{\mathrm{P,top}}}}
\newcommand{\COBOLD}{{\sc CO$^5$BOLD}}

\newcommand{\changed}{}

\newcommand{\I}{\ensuremath{I}}
\newcommand{\Irot}{\ensuremath{\tilde{I}}}
\newcommand{\F}{\ensuremath{F}}
\newcommand{\Frot}{\ensuremath{\tilde{F}}}
\newcommand{\vsini}{\ensuremath{V\sin(i)}}
\newcommand{\vvsini}{\ensuremath{V^2\sin^2(i)}}
\newcommand{\vsinimu}{\ensuremath{\tilde{v}}}
\newcommand{\rotint}{\ensuremath{\int^{+\vsinimu}_{-\vsinimu}\!\!d\xi\,}}
\newcommand{\imu}{\ensuremath{m}}
\newcommand{\imupone}{\ensuremath{{m+1}}}
\newcommand{\nmu}{\ensuremath{N_\mu}}
\newcommand{\msum}[1]{\ensuremath{\sum_{#1=1}^{\nmu}}}
\newcommand{\wmu}{\ensuremath{w_\imu}}

\newcommand{\tchar}{\ensuremath{t_\mathrm{c}}}
\newcommand{\Nt}{\ensuremath{N_\mathrm{t}}}

\title{The photospheric solar oxygen project:}
\subtitle{III. Investigation of the centre-to-limb variation of the 630\,nm [O\,I]-{Ni}\,{I} blend}

\author{
E. Caffau      \inst{1,2}\and
H.-G. Ludwig   \inst{2,1}\and
M. Steffen     \inst{3,1}\and
W. Livingston  \inst{4}\and
P. Bonifacio   \inst{1}\and
J.-M. Malherbe \inst{5}\and
H.-P. Doerr    \inst{6,7}\and
W. Schmidt     \inst{6}
}

\institute{
GEPI, Observatoire de Paris, PSL Resarch University, CNRS,
Univ Paris Diderot, Sorbonne Paris Cit\'e, Place Jules Janssen, 92195
Meudon, France
\and
Zentrum f\"ur Astronomie der Universit\"at Heidelberg, Landessternwarte, 
K\"onigstuhl 12, 69117 Heidelberg, Germany
\and
Leibniz-Institut f\"ur Astrophysik Potsdam, An der Sternwarte 16, 
D-14482 Potsdam, Germany
\and
National Solar Observatory, Tucson, AZ 85726
\and
LESIA, Observatoire de Paris, PSL Resarch University, CNRS,
Univ Pierre et Marie Curie, Univ Paris Diderot, Sorbonne Paris Cit\'e, Sorbonne Univertit\'es, 
Place Jules Janssen, 92195 Meudon Cedex, France
\and
Kiepenheuer-Institut f\"ur Sonnenphysik, Sch\"oneckstra{\ss}e 6, 79104 Freiburg, Germany 
\and
Max-Planck-Institut for Solar System Research, Justus-von-Liebig-Weg 3, 37077 G\"ottingen, Germany
}
\authorrunning{Caffau et al.}
\titlerunning{Solar oxygen project III}
\offprints{}
\date{Received ...; Accepted ...}

\abstract
{The solar photospheric abundance of oxygen is still a matter of 
debate. For about ten years some determinations have favoured  
a low oxygen abundance which is at variance with the value inferred
by helioseismology.
Among the oxygen abundance indicators,
the forbidden line at 630\pun{nm} has often been considered  the 
most reliable even though it is blended with a \ion{Ni}{i} line.
In Papers~I and Paper~II of this series we reported a discrepancy in the
oxygen abundance derived from the 630\pun{nm} and the subordinate [O\,I]
line at 636\pun{nm} in dwarf stars, including the Sun.}
{Here we analyse several, in part new, solar observations of the
the centre-to-limb variation of the spectral region including the 
blend at 630\pun{nm} in order to separate the individual contributions of 
oxygen and nickel.}
{We analyse intensity  spectra observed at different
limb angles in comparison with
line formation computations performed
on a CO5BOLD 3D hydrodynamical simulation of the solar atmosphere.}
{The  oxygen abundances obtained from
the forbidden line at different limb angles are inconsistent if the commonly 
adopted nickel abundance of $6.25$ is assumed in our local thermodynamic equilibrium computations.
With a slightly lower nickel abundance, A(Ni) $\approx 6.1$, we obtain 
consistent fits indicating an oxygen abundance of A(O)=$8.73\pm 0.05$. At 
this value the discrepancy with the subordinate oxygen line remains.}
{The derived value of the oxygen abundance supports the notion of a rather low
  oxygen abundance in the solar photosphere. However, it is disconcerting
that the forbidden oxygen lines at 630 and 636\pun{nm} give noticeably different
results, and that the nickel abundance  derived here from the 630\pun{nm} 
blend is lower than expected from other nickel lines.}

\maketitle


\section{Introduction}

Although oxygen is the third most abundant element in the Sun 
(after hydrogen and helium), its precise abundance is currently 
under discussion. Very early on, \citet{sedlmayr74} derived a rather 
low oxygen abundance in comparison to the accepted value at the time, 
and attributed the difference to
the important role of departures from local thermodynamic equilibrium (LTE) for the oxygen 
infrared triplet at 777\pun{nm}. The debate gained momentum thanks to the pioneering works of 
\citet{kiselman95}, \citet{ALA01}, \citet{hhoxy}, and \citet{asplund04} who
were the first to apply hydrodynamical simulations (hereafter 3D models)
to the determination of the solar oxygen abundance.
Their results advocated a downward revision of the solar
oxygen abundance, and therefore of the solar metallicity $Z$,
which now appears to be in conflict with
the  helioseismic data \citep[see][for a review of the topic]{basu}.
In the last ten years much effort has been dedicated to confirming or dismissing
the downward revision of the solar oxygen abundance \citep{asplund04,asplund09}.
\citet{pereira09} analysed the [OI] line; they put A(Ni)=6.22 and derived 
A(O)=8.64 from their centre-disc spectrum.
\citet{ayres06} investigated the infrared lines of CO while
\citet{ayres08} analysed the [OI] line at 630\,nm, both indicating a ``high'' 
oxygen abundance.
 \citet{centeno08} also analysed the [OI] line at 630\,nm and derived a high value,
${\rm A(O)}=8.86\pm 0.07$.
\citet{socasnavarro07} analysed the oxygen triplet deriving a ``low'' NLTE oxygen abundance.
\citet{socasnavarro14} investigated the [OI] line with the support of an empirical
three dimensional model, obtaining a high A(O), but with a large
uncertainty and with
the possibility of deriving a low oxygen abundance by slightly changing the model.
This paper is particularly interesting because it shows the limiting factors of the models we use.

Our project aims at investigating the oxygen indicators in the solar spectrum 
and deriving the solar oxygen abundance based on a \cobold\ 3D model of the 
solar atmosphere. In a previous paper of this series \citep{oxy}, hereafter Paper~I, we derived a measure of the solar oxygen abundance from 
an ensemble of atomic lines. Among the lines employed in that investigation
the two [OI] lines at $\lambda =630$\pun{nm} and $636$\pun{nm}
were singled out as potentially problematic.
The abundance we derived from the two lines 
differed by 0.1\,dex, both when using the 3D model and
when using the 1D Holweger-M\"uller  solar
model (\citealt{hhsunmod, hmsunmod}, hereafter HM).
This is certainly a good reason to  investigate these lines further.
In addition, the [OI] lines, and in particular the stronger 630\pun{nm}
line, have often been considered  the best indicators
of the oxygen abundance, given that they are weak
and formed in conditions of  LTE.
When using other atomic lines to determine the solar oxygen 
abundance, one is forced to introduce an additional
uncertainty related to the computation of the NLTE corrections,
which stems from our lack of knowledge of the efficiency of 
excitation and ionisation of oxygen by collisions with hydrogen atoms.
Only for a few atoms do we know the cross sections of inelastic collisions with
neutral hydrogen from either experiments or quantum-mechanical computations.
The use of the Drawin approximation \citep{drawin69,steenbock84} has been 
seriously criticised by \citet{barklem11}.
For further discussion on this point see also \citet{oxyiv}.
We estimated the abundance uncertainty due to this
unknown parameter as about $0.03$~dex (Paper~I).
If one could reliably determine the oxygen abundance
from the forbidden lines, the role of hydrogen 
collisions could be constrained by requiring that
forbidden and permitted lines yield the same abundance.

We investigated in \citet{oxy13} (Paper~II) the forbidden lines in a sample 
of stars from high-quality  spectra and concluded that the disagreement in the
oxygen abundance derived from the two [OI] lines is generally
evident in dwarf stars, while it vanishes in giants.

In the present investigation we exploit observations of solar 
intensity spectra at different limb angles to put additional constraints on 
the abundance analysis of the forbidden oxygen line at 630\,nm. 


\section{Atomic data and model atmospheres}

For the forbidden oxygen line and for the \ion{Ni}{i} line we employ
the same atomic data as used in Paper~I. 
The adopted $f-$values are from \citet{SZ}.
The damping parameters have no influence on the profile of these weak lines.

Our analysis is based on a 3D model atmosphere
computed with the \cobold\ code \citep{freytag12}. 
The box-size of the \cobold\ 3D solar model atmosphere is 
$5.6\times 5.6\times 2.25$\,Mm, with a resolution of $140\times 140\times 150$. 
The simulation covers a range in time 
of 2.1\pun{h}, represented by 20 snapshots;
this time interval corresponds to 25 periods of the solar 5\,min oscillations 
and about 18 convective turn-over times. The snapshots were taken at equal time intervals and were selected to 
provide a good representation of the statistical properties of the complete 
temporal sequence obtained in the \cobold\ run.
The model has been described in \citet{solarphy} and used for the chemical investigation
of the photospheric solar abundance of several elements.
The spectrum synthesis code employed is 
\linfor\footnote{http://www.aip.de/Members/msteffen/linfor3d/linfor\_3D\_manual.pdf}.
The spectra are calculated for the same $\mu$ inclinations as the observed spectra,
and for four values of azimuthal angles.
The spectrum corresponding to each $\mu$-value is an average over all surface elements,
all azimuthal directions, and all time snapshots.


\section{Observations}

We analysed five sets observations from
\begin{itemize}
\item
\textit{Hinode} (five disc positions);
\item
the Swedish Solar Telescope, SST (four disc positions);
\item
the McMath-Pierce Solar Telescope, Kitt Peak (six disc positions);
\item
the Vacuum Tower Telescope, VTT (five disc positions);
\item
THEMIS (eight disc positions).
\end{itemize}
We think that all of these observations are extremely
useful to the discussion of the solar oxygen problem. However, two sets
clearly stand out for different reasons:
the \textit{Hinode} observations, taken from space, are not contaminated by
any telluric feature; the VTT data are of higher spectral resolution than
any other data set.
The Kitt Peak and the SST observations are of similar quality,
although the latter ones have higher signal-to-noise ratios.
The THEMIS observations present some difficulty in the determination of 
the continuum. In the following, each set of observations is discussed in 
a dedicated section.

We restrict our analysis to the spectra at $\mu > 0.2$ for all observed sets
of spectra for the following reasons. 
At smaller $\mu$-values sphericity effects become large enough to
influence the computed line shapes.
The most extreme limb spectra are formed in more external regions of the solar photosphere,
and more so for oxygen than for nickel (see Figs. \,\ref{contf} and \ref{contfNi}, respectively). 
The thermal structure and velocity fields of these
external low-density layers of the atmosphere are
more difficult to model correctly, since they are more
sensitive  to the upper boundary condition of the 
simulation box.

\subsection{\textit{Hinode} data}

\textit{Hinode} (JAXA/NASA) observations were obtained with the Solar
Optical Telescope (SOT, 50\,cm aperture) and the Spectro Polarimeter (SP)
using a thin slit of 0.3\,arcsec $\times$ 160\,arcsec.
Details of the solar granulation
are clearly visible (granules, intergranules). Observations
were performed at ten locations along the solar
equator with the slit orthogonal to the equator direction
so that pixels along the slit are seen at a mostly constant heliocentric 
direction, cosine~$\mu$.
In fact, there is a small $\mu$ variation
along the slit because pixels at the end of the slit ($\pm 80^{\prime\prime}$)
do not exactly have the $\mu$ of the slit centre,
but it is a small effect.
The resolution provided by the
spectrograph is about 200\,000 so that the spectra
exhibit resolved features of 3\,pm at 630\,nm. The sampling is 2\,pm.
The SP is usually centred on the \ion{Fe}{i} 630.15 and 630.25\,nm lines,
but it is possible to move  the spectral coverage slightly 
to [O\,I] 630.03\,nm and \ion{Fe}{i} 630.15\,nm.
The CCD provided spectral images of 112 $\times$ 512 pixels
corresponding to a spectral/spatial field of 0.23\,nm $\times$ 160\,arcsec. Individual images
have a signal-to-noise (S/N) ratio of 1000 in the continuum.
For each position along the equator, a small scan along 24\,arcsec
was made in the equatorial direction, which  is the reason why we selected only
the central part of the scan to stay at constant $\mu$ along the equator. 
Two sides of the CCD were exposed (1.6\,s exposure time) and we used both.
All Stokes parameters were recorded (full Stokes mode I, Q, U, V), but
we selected only intensities, I. However, the full Stokes mode allowed us
to check on the \ion{Fe}{i} 630.15\,nm line (Lande factor 1.67) that only a few pixels 
were affected by magnetic fields of the solar network.
The curvature of lines was corrected by
a parabolic fit derived from the strong \ion{Fe}{i} 630.15\,nm line close to
the [O\,I] 630.03\,nm line with flat-field data.
Then, a summation was done in the spatial direction
along the slit to get high precision spectra for
various $\mu$-values with a final nominal
S/N ratio of 20\,000 in the continuum. 

\subsection{SST data}

These spectra were observed at the Swedish Solar Telescope in May 2007.
They are described in \citet{pereira} and analysed in \citet{pereira09}.
The observations were taken with slit width of 0.11\,arcsec, the spectral 
resolving power is 200\,000, the four $\mu$-values used are 
0.999, 0.793, 0.608, and 0.424.

\subsection{McMath-Pierce data (WCL)} 

The observational data consists of solar intensity spectra for nine 
heliocentric angles $\theta$  
($\mu = 1.00,~0.87,~0.66,~0.48,~0.35,~0.25,~0.17,~0.12,~0.10$), 
observed with the McMath-Pierce Solar Telescope at Kitt Peak,
using the rapid scan double-pass
spectrometer \citep[see][and references therein]{Livingston}
in single-pass mode.  
The observations were taken on September 14,  2006; the local time was about 09h30m to 10h30m.
The resolution was set by the slit width: 0.06\pun{mm} entrance, 0.1\pun{mm} exit.
This yields a spectral resolving power of about 90\,000.
The scan length, in wavelength, is 2048 points.
The image scale is 2\farcs{5}\pun{mm$^{-1}$}.
The slit was set parallel to the nearest limb.

The setting was fixed in order to observe
the two oxygen forbidden lines at 630\pun{nm} and 636\pun{nm}.
The observations were performed on the north-south axis in the sky,
from solar centre to limb. For each line we obtained two spectra for any 
$\mu$-value off-centre,
and four spectra of the intensity at disc-centre.
Each observation freezes the solar five-minute oscillations, 
covering a shorter interval of time.
Effective exposure time was 
100\pun{s} for $\mu \ge 0.25$ (ten scans of 10\pun{s}), and
200\pun{s} for $\mu \le 0.17$ (twenty scans of 10\pun{s}).
We summed the spectra at equal $\mu$ to improve the signal-to-noise ratio.
No wavelength calibration source was observed and we used
the available telluric lines to perform the wavelength 
calibration. The positions of the telluric lines were
measured on the
high-resolution, high signal-to-noise ratio, solar spectra of the disc-centre atlas of \citet{neckelobs}, 
assuming a linear dispersion relation.
We will refer to this set of observations as WCL  data.

\subsection{VTT data}

The VTT data were obtained on Tenerife, Canary Islands, with the new prototype instrument LARS
(Lars is an Absolute Reference Spectrograph). The set-up is based on the
high-resolution vertical Echelle spectrograph of the VTT and employs a laser
frequency comb (LFC) wavelength calibration source (see
e.g. \citealt{Murphy2007}, \citealt{Wilken.etal.2012}, \citealt{steinmetz08}) and a single-mode fibre
feed (SMFF). The wavelength calibration obtained from the LFC is accurate to
1\,m\,s$^{-1}$. Light from the VTT focal plane is coupled to the fibre from
a circular field of view (FOV) with a diameter of 10\,arcsec on the
sky. Flat fields are obtained with a fibre-coupled continuum lamp, resulting
in an excellent gain-table correction. The set-up provides a spectral purity
and fidelity that is, to the best of our knowledge, not attainable with any
other instrument currently available for solar observations.
An early version of the instrument is described by \citet{doerr.etal.2012,doerr.etal.2012b} and by \citet{probst15}.

The spectra used in this work were obtained during several campaigns in the
testing and commissioning phase of the instrument in August and December
2013 and in March and July 2014. Repeated observations were carried out at
disc-centre and at heliocentric angles corresponding to $\mu = \left\{0.8,
0.6, 0.4, 0.3, 0.2\right\}$, but only the data for $\mu = 1.0\ldots 0.3$ were used.
Individual observations ran between 15 and 30 minutes,
depending on the observing conditions. Seeing usually was moderate to bad
which is, however, not important as the instrument was fed with the spatial
average of a 10\,arcsec FOV in all cases. We took some care to avoid regions with
magnetic activity by monitoring G-Band context images and Calcium\,II K
full-disc images provided by the ChroTel telescope in real-time
\citep{kentischer.etal.2008}. The absolute pointing accuracy of the VTT is
better than~5\,arcsec and was checked frequently and adjusted if necessary.

All spectra were corrected for dark-current, and the gain-table obtained from
the continuum lamp was applied. Residual spectral slopes were corrected by
fitting a second-order polynomial to selected high-points in our spectra and
to the corresponding points in the Kitt Peak Fourier Transform Spectrometer (FTS) disc-centre spectrum as
provided by \cite{Neckel1999}. The ratio of both polynomials gave the
correction factor that was applied to our spectra to force their continuum
to be identical with that of the FTS spectrum. The wavelength calibration
was derived from the LFC calibration spectra. The laser lines are completely
unresolved by the spectrograph and can be used to measure its instrumental
profile (IP). From the full width at half maximum of the IP, we derived a
spectral resolution between 600\,000 and 750\,000 for these observations,
depending somewhat on the alignment of the SMFF, which was changed between the
individual observing campaigns.

The calibrated spectra were interpolated to an equidistant wavelength scale
of~1\,m\AA{}\,pixel$^{-1}$ (native sampling is 2.73\,m\AA{}
pixel$^{-1}$). The LFC calibration naturally removes the effect of any
instrumental drift and the radial velocity of each observed spectrum was
subtracted, resulting in a wavelength scale that is defined in the solar
reference frame. All spectra measured at the same heliocentric angle were
then summed to increase the S/N ratio. The S/N ratio was measured in the
continuum at 6300\,\AA{} and it reaches a value of~6900 for the disc-centre
spectra.  Table\,\ref{vttdata} provides a census of the calibrated spectra
that are used in this work.

\begin{table}
  \caption{Census of the VTT data. Data from several individual observations
    is summed  for each heliocentric angle, corresponding to total
    observation times between 36 and 228\,minutes.}
  \label{vttdata}
  \centering
    \begin{tabular}{lrrr}
      \hline
      \noalign{\smallskip}
      $\mu$ & ${\rm N}_{\rm obs}$ & ${\rm T}_{\rm obs}$ (min) & S/N \\
      \hline
      \noalign{\smallskip}
      1.0 & 8 & 228.4 & 6900\\
      0.8 & 2 & 49.9  & 4600\\
      0.6 & 5 & 127.6  & 6400\\
      0.4 & 2 & 35.8  & 3200\\
      0.3 & 2 & 53.6 & 4400 \\
      \hline
    \end{tabular}
\end{table}

\subsection{THEMIS data}

THEMIS observations were obtained with the  eight-metre spectrograph in classical spectroscopic mode
using a thin slit of 0.5\,arcsec $\times$ 120\,arcsec.
However, the seeing was never better than 1 to 2 arcsec,
so that details of the solar granulation
are not visible. Several scans along the solar
radius were performed with the slit orthogonal
to the radius direction (so that pixels along the slit
are seen at almost constant $\mu$). The interference
order was 36 and was selected by the predisperser
spectrograph. The final resolution provided by the
eight-metre spectrograph is about 300\,000 so that the spectra
exhibit resolved features of 2\,pm at 630\,nm.
The CCD provides spectral images of 512 x 512 pixels
corresponding to a spectral/spatial field of 0.5 nm x 120 arcsec with
exposure time of 100 ms. Individual images
have a S/N ratio of 250 in the continuum.
For each position of the slit on the sun, 360 images were recorded
and summed to improve the S/N ratio to
about 5000. The curvature of lines was corrected by
a parabolic fit derived from the strong \ion{Fe}{i} 630.15\,nm line close to
the [O\,I] 630\,nm line with flat-field data.
Then, a summation was done in the spatial direction
of the slit to get high-precision spectra for
various $\mu$-values with
a final S/N ratio of 100\,000 in the continuum.

Eight positions on the solar disc are observed, at
$\mu = \left\{1.00, 0.96, 0.88, 0.72, 0.66, 0.48, 0.35, 0.25\right\}$.

\section{Data analysis}

In our previous analysis of the solar oxygen abundance based on high-resolution, 
high S/N, central intensity and flux spectra (Paper~I), 
we found that the oxygen abundance A(O) from the 630\pun{nm} line
is 0.1\,dex lower than the value obtained from the 636\pun{nm} line.
It is known that the 630\pun{nm} line is blended with a \ion{Ni}{i}
line, whose \loggf\ has been well determined \citep{Johansson}.
On the other hand the 636\pun{nm} line, which is weaker than the
other [OI] line, lies on the red wing of a
\ion{Ca}{i} autoionisation line, and is blended with CN molecular
lines, whose \loggf\ are more uncertain.
From these considerations one should give preference to the A(O)
determination from the 630\pun{nm} line,
and for this reason we concentrate on this line in the present work.

\begin{figure}
\resizebox{\hsize}{!}{\includegraphics[clip=true,angle=0]{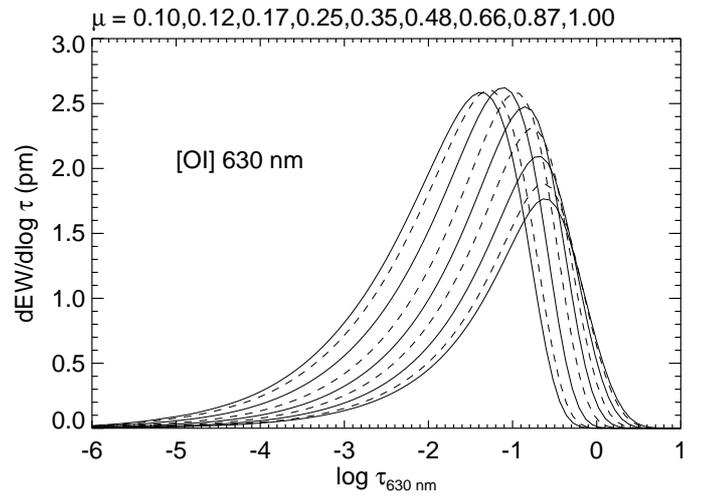}}
\caption{The contribution function of the EW of the pure [OI] 630\pun{nm} 
  line for various $\mu$-angles as a function of monochromatic optical depth 
  in the continuum. The $\mu$-values are given at the top of the panel, 
  the curves are plotted alternating as solid and dashed lines. The ones 
  peaking at higher optical depth correspond to higher $\mu$-values.}
\label{contf}
\end{figure}

\begin{figure}
\resizebox{\hsize}{!}{\includegraphics[clip=true,angle=0]{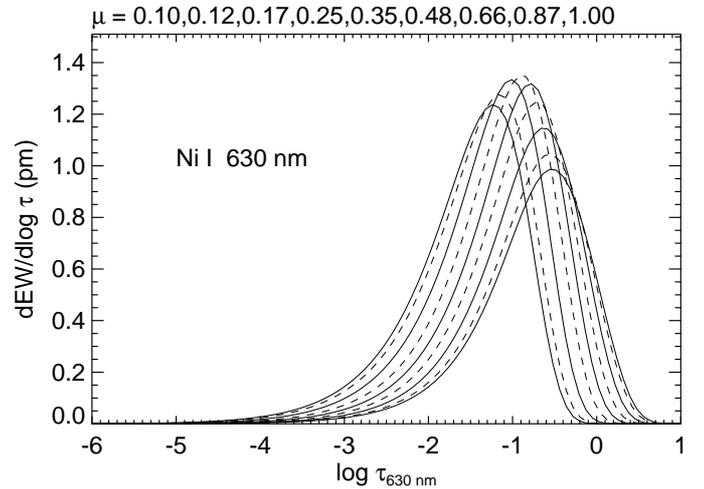}}
\caption{Like Fig.\,\ref{contf} but for the blending \ion{Ni}{i} line at 630\pun{nm}.}
\label{contfNi}
\end{figure}

The two forbidden lines are very similar from the point of view of atomic data,
and are formed in the same atmospheric layers.
Based on the 3D model, we computed the contribution function 
for the equivalent width (EW) of the [OI] 630 nm line
at the various $\mu$-values of the Kitt Peak intensity spectra
with the spectral synthesis code Linfor3D.
These are shown in Fig.\,\ref{contf}.
As $\mu$ decreases, the contribution function becomes broader
and the peak moves to higher photospheric layers.
For $\mu =1.0$ the maximum value of the contribution function
is at $\log\tau_{\rm 630~nm} = -0.6$ and has a full width at half maximum (FWHM) of 1.2
in $\log\tau$; for $\mu =0.25$ the maximum is at --1.0 and
the FWHM 1.5 \relax in $\log\tau$.
The spectra at different limb angles sample slightly different thermodynamic 
conditions at which the 630\pun{nm} blend is formed. In principle, different
spectral lines probe different atmospheric layers, and show a different 
centre-to-limb variation. This can help to disentangle 
the contribution of oxygen and nickel to the the total equivalent width of the
blend. In Fig.\,\ref{contfNi}, the contribution function of the EW is shown
for the Ni contribution. The similarity of the $\mu$-dependence of the 
contribution functions of oxygen (Fig.\,\ref{contf}) and nickel 
(Fig.\,\ref{contfNi}) already indicates that it is still a difficult task.

\begin{figure}
\resizebox{\hsize}{!}{\includegraphics[clip=true,angle=0]{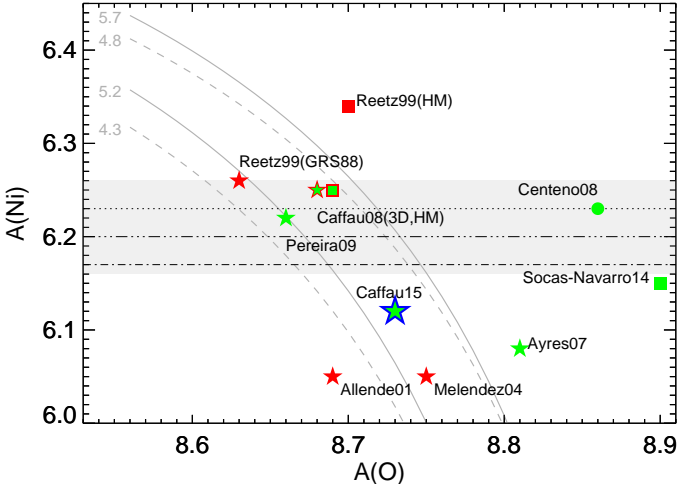}}
\caption{The plot shows the oxygen and nickel abundances derived in the past years.
The grey bar is the presently discussed range of the solar Ni abundance. 
Red symbols mean disc-integrated spectrum, green symbols
disc centre. Stars indicate theoretical models, squares the HM model. 
The round dot of Centeno08: in the work a sun-spot spectrum is used.
The dashed lines are lines of constant EW for disc centre, solid for
disc-integrated light. This figure is an updated version of the one shown by \citet{hgl10}.
The big green-blue star represents the result of this investigation.
The horizontal dotted line at A(Ni)=6.23 represents values from \citet{grevesse98}; 
the horizontal dot-dashed line at A(Ni)=6.17 the value from \citet{scott09};
the horizontal dot-dashed line at A(Ni)=6.20 the value from \citet{scott14}.
}
\label{aoniplot}
\end{figure}

The feature we investigate is a O-Ni blend. The observed spectra at disc-centre agree in that the EW of the blend is of the order of 0.47\,pm. It is 
clear that the derived abundances of oxygen and nickel are related: any 
increase in A(O) implies a decrease in A(Ni) and vice-versa. 
A high oxygen abundance has to be associated with a low Ni 
abundance (a high A(O) implies a high contribution to the EW of the blend due to oxygen so that
the fraction of EW left for Ni is small, implying a low A(Ni)) and a low A(O) to a high A(Ni). 
The situation is 
summarised in Fig.\,\ref{aoniplot},
where we present the relative abundances of oxygen and nickel for a given
range in EW based on 3D analysis.

\subsection{Line profile fitting}

We determine the oxygen abundance of the [OI] line at 630\,nm from line profile fitting.
In this way we take into account the line strength and the line shape at the same time.
As in \citet{fit03}, we fitted the line profile interpolating in a
grid of synthetic spectra. To use the complete information, we  
simultaneously fitted all the limb angles of the solar intensity profiles, discarding
the spectra closest to the limb.
The same method is also used to analyse the \ion{O}{i} triplet lines and is extensively explained in \citet{oxyiv}.
The previous analysis of this [OI] line by \citet{pereira09} focused only on the centre-disc spectrum
for the line profile fitting procedure.

To fit the observed spectra with the grid of 3D spectra, we must
normalise the wings of the [OI] line to unity because 
the 3D synthetic spectra contain only the lines of oxygen and nickel.
No molecular or other line contributions that would prevent normalised spectra
from going to unity are considered in the spectrum synthesis.
To try to be as objective as possible, and to avoid
a subjective normalisation of the spectra, we select two 
ranges that are not contaminated by lines on the red and on the blue side of the
[OI] line, and use these two ranges to {\it \emph{pseudo-normalise}} the 
spectra at all $\mu$-angles.
Using this procedure, there are between 10 and 18 free fitting parameters
(A(O), A(Ni), and wavelength shift and continuum level for each of the 
spectra, observed at four to eight $\mu$-values).
We find that the fits are not always very robust.
The minimum of the $\chi ^2$ is not narrow and in the region of the best A(O) and A(Ni) values 
there are several local minima.
By changing the input parameters,
the values we derive for A(O) and A(Ni) are not always the same but fall in a limited range 
spanning $\la 0.02$\,dex.
We restricted the fit to reasonable values
of oxygen abundance ($8.6\le {\rm A(O)}\le 8.85$) and nickel abundance ($5.85\le {\rm A(Ni)}\le 6.35$).

\begin{figure}
\resizebox{\hsize}{!}{\includegraphics[clip=true,angle=0]{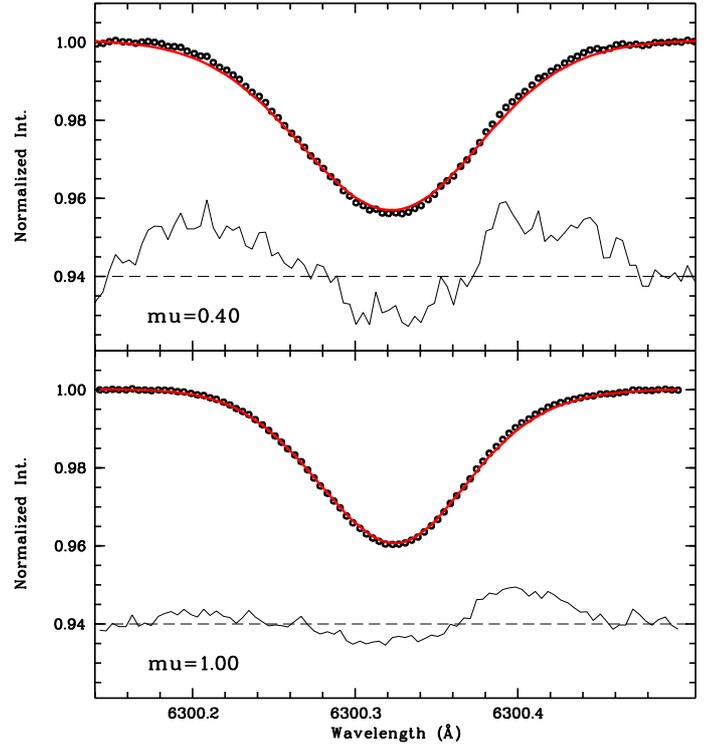}}
\caption{ VTT observations with a resolving power of 700000 
(black circles)  compared to the 3D synthetic profile (solid red) at
$\mu=0.4$ (top) and at disc-centre (bottom).
Below each profile the residual ($\times 10$, displaced by $+0.94$) is shown.}
\label{vtt100}
\end{figure}

Another problem that needs to be mentioned is that we did not apply any 
instrumental broadening to the synthetic profile (which would correspond to about 0.5\,km/s for
the VTT data and 1.5\,km/s for \textit{Hinode} spectrum) because it is already
as broad as or slightly broader than the observed one even without any 
extra broadening. The problem is visible in Fig.\,\ref{vtt100} where two 
VTT profiles, observed at centre-disc and at $\mu =0.4$, respectively,
are compared to a synthetic profile with the oxygen and nickel abundances
of A(O)=8.72 and A(Ni)=6.17.
From the figure it is clear that the agreement is worse towards the limb. 

The five sets of observations provide A(O) and A(Ni) in limited ranges. 
The results are summarised in Table\,\ref{anires}.
When we average the results we obtain ${\rm A}\left({\rm O}\right)=8.735\pm 0.017$
and ${\rm A}\left({\rm Ni}\right)=6.096\pm 0.049$.
If we restrict our average to the data of \textit{Hinode}, VTT, and SST,
which we think are of higher quality, we obtain:
${\rm A}\left({\rm O}\right)=8.725\pm 0.012$ and ${\rm A}\left({\rm
  Ni}\right)=6.122\pm 0.034$,
 the uncertainty being the observed-to-observed scatter.
This latter value for the Ni abundance is in agreement with the 
value of ${\rm A(Ni)} = 6.17\pm 0.02$ from \citet{scott09} and in 
reasonable agreement with ${\rm A(Ni)}=6.20\pm 0.04$ from \citet{scott14},
who used an improved 3D solar model atmosphere
with respect to \citet{scott09}. Our nickel abundance is almost
within one $\sigma$ of the \citet{scott09} and of the \citet{scott14} results.

\begin{table}
\caption{Summary of the results of the line-profile fitting.}
\label{anires}
\begin{center}
{
\begin{tabular}{lrr}
\hline
\noalign{\smallskip}
Spectrum & A(O) & A(Ni)\\
\hline
\noalign{\smallskip}
\textit{Hinode} & $8.71$ & $6.16$ \\
SST             & $8.73$ & $6.13$ \\
WCL             & $8.74$ & $6.10$ \\
VTT             & $8.74$ & $6.08$ \\
THEMIS          & $8.76$ & $6.02$ \\
\noalign{\smallskip}
\hline
\end{tabular}
}
\end{center}
\end{table}

\subsubsection{\textit{Hinode} data}

\begin{figure}
\resizebox{\hsize}{!}{\includegraphics[clip=true,angle=0]{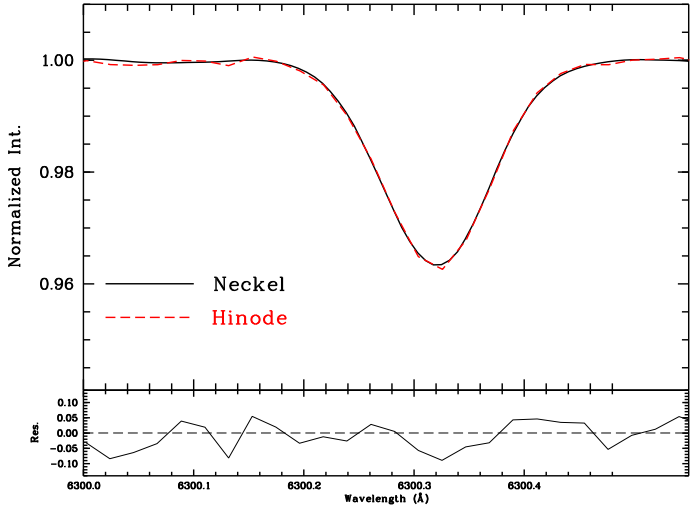}}
\caption{Comparison of the \textit{Hinode} centre-disc observation
(dashed red) with the Neckel centre-disc spectrum (solid black); the agreement 
is very good. The Neckel spectrum has been broadened by 2\,km/s for comparison with  the
\textit{Hinode} observation.
In the lower panel the residual ($\times 100$) is shown.}
\label{comHNI}
\end{figure}

We investigated the possibility of scattered light, but
it does not seem to be a problem for the \textit{Hinode} observations.
The comparison of the \textit{Hinode} centre-disc spectrum with the Neckel 
centre-disc data gives a perfect agreement (see Fig.\,\ref{comHNI}).

\begin{figure}
\resizebox{\hsize}{!}{\includegraphics[clip=true,angle=0]{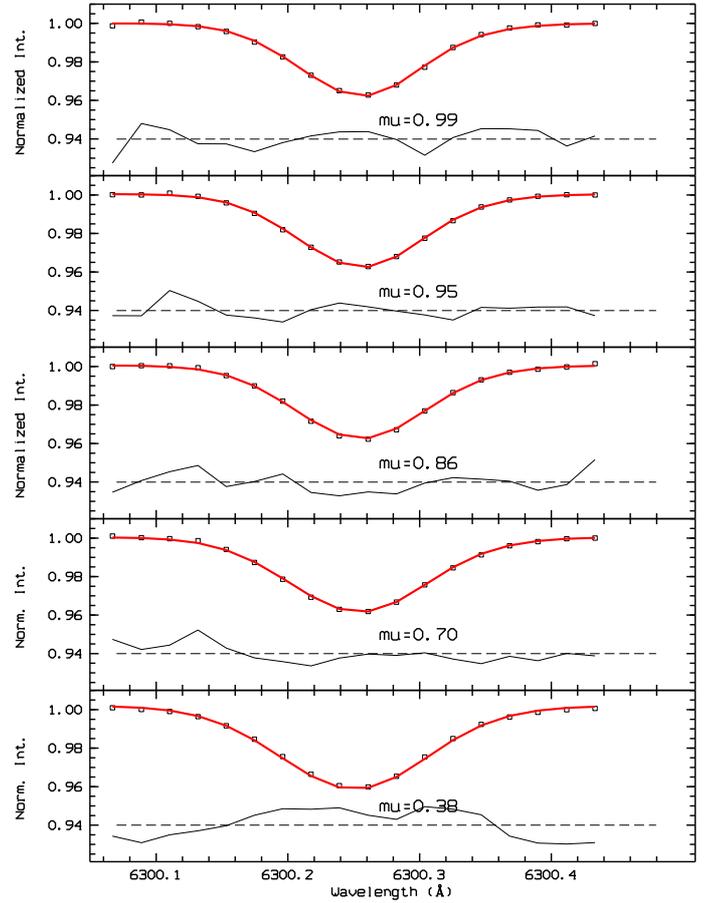}}
\caption{Fitted 3D line profiles for the five $\mu$-values (red)
 compared to the \textit{Hinode} observed spectra (black).
 From the simultaneous fit we derive A(O)=8.71 and A(Ni)=6.16.
Below each profile the residual ($\times 10$, displaced by +0.94) is shown.
}
\label{plotfithinode}
\end{figure}

We fitted the five observed spectra corresponding to $\mu$-values of 
0.99, 0.95, 0.86, 0.70, and 0.38 for two sets of \textit{Hinode} data. 
The best fit of the five $\mu$-values gave A(O)=8.71 and A(Ni)=6.16 
(see Fig.\,\ref{plotfithinode}).

\subsubsection{SST data}

\begin{figure}
\resizebox{\hsize}{!}{\includegraphics[clip=true,angle=0]{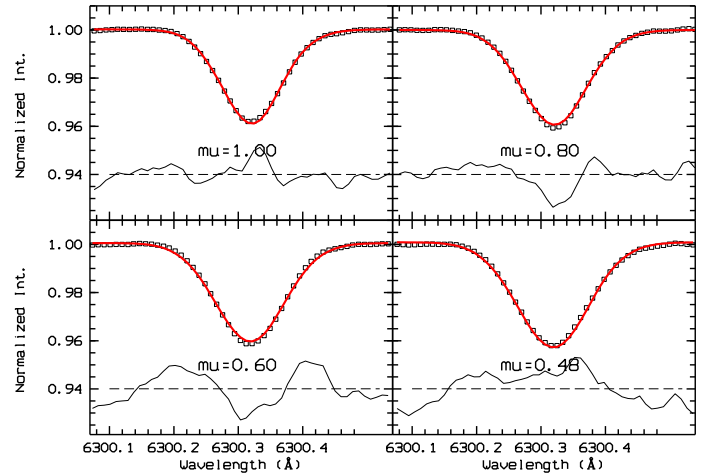}}
\caption{Fitted 3D line profiles for the four $\mu$-values (red)
 compared to the SST observed spectra (black).
 From the simultaneous fit we derive A(O)=8.73 and A(Ni)=6.13.
Below each profile the residual ($\times 10$, displaced by +0.94) is shown.
}
\label{plotfitpereira}
\end{figure}

The best fit of the four $\mu$-values gave A(O)=8.73 and A(Ni)=6.13 
(see Fig.\,\ref{plotfitpereira}). 
The fit looks good and the result is in agreement with the values derived from the other observed spectra.
The spectra show no signs of scattered light.

\subsubsection{WCL data}

\begin{figure}
\resizebox{\hsize}{!}{\includegraphics[clip=true,angle=0]{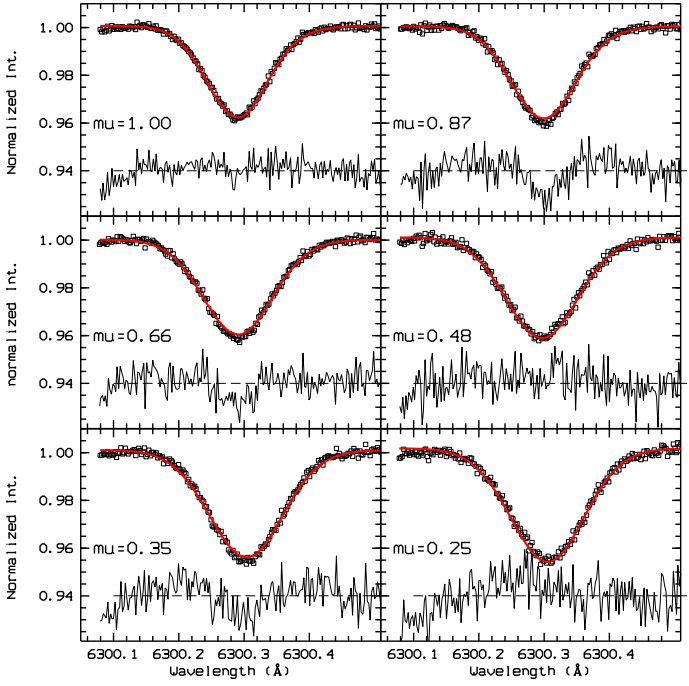}}
\caption{Fitted 3D line profiles for the six $\mu$-values (red)
 compared to the WCL observed spectra (black).
 From the fit we derive A(O)=8.74 and A(Ni)=6.10.
Below each profile the residual ($\times 5$, displaced by +0.94) is shown.
}
\label{plotfitbl}
\end{figure}

We compared the WCL observed spectra at $\mu = 1.0$ to the solar atlases of
\citet{delbouille} and \citet{neckelobs} and discovered that our WCL
disc-centre spectrum has a smaller EW  (by roughly 4\%) than in both high-resolution solar
atlases.  This could be due to scattered light in the 
spectrograph, solar activity, or some other unknown effect.  
We corrected the data by this factor, assuming that the scattered light
affects the spectra at all $\mu$-angles in the same way.
We restricted the line profile fitting to the six spectra with $\mu > 0.2$.
Using this procedure, we fitted the observed spectra with 14 free parameters
(A(O), A(Ni), and wavelength shift and continuum level for each of 
the six spectra) and obtained A(O)=8.74 and A(Ni)=6.10 
(see Fig.\,\ref{plotfitbl}).

\subsubsection{VTT data} \label{vttanalysis}

We find no evidence for a contamination by scattered light in these spectra.

\begin{figure}
\resizebox{\hsize}{!}{\includegraphics[clip=true,angle=0]{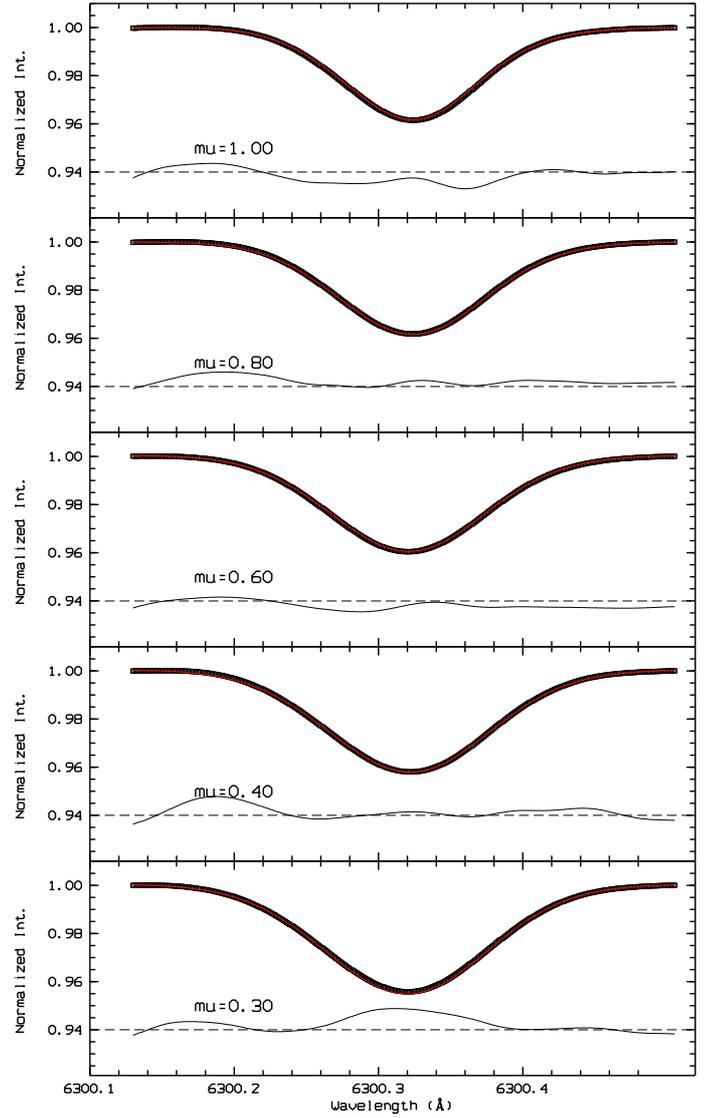}}
\caption{Fitted 3D line profiles for five $\mu$-values (red)
 compared to the VTT observed spectra (black). From this fit we derive
A(O)=8.72 and A(Ni)=6.12.
Below each profile the residual ($\times 10$, displaced by $+0.94$) is shown.}
\label{plotvttbro2}
\end{figure}

We note that we are not able to fit the high-quality VTT spectra 
because our synthetic profiles are too broad. We therefore decided to 
broaden the observed profiles and we were able to obtain a very good fit 
(see  Fig.\,\ref{plotvttbro2} for an example).
The required extra broadening is $\mu$-dependent:
1.3, 1.7, 1.8, and 1.9\kms for $\mu$ positions 
1.0, 0.8, 0.6, and 0.4, respectively. In this way we derive
A(O)=8.74 and A(Ni)=6.08.
With five $\mu$-values, the most extreme limb component, $\mu =0.3$, required an extra broadening of 1.9\kms.
In this case we derive A(O)=8.72 and A(Ni)=6.12 (see Fig.\,\ref{plotvttbro2}).
The oxygen and nickel abundances we derive in the two cases are very similar,
but we prefer to use the case with four $\mu$-values because the
agreement with the observed data is better.

\subsubsection{THEMIS data}

\begin{figure}
\resizebox{\hsize}{!}{\includegraphics[clip=true,angle=0]{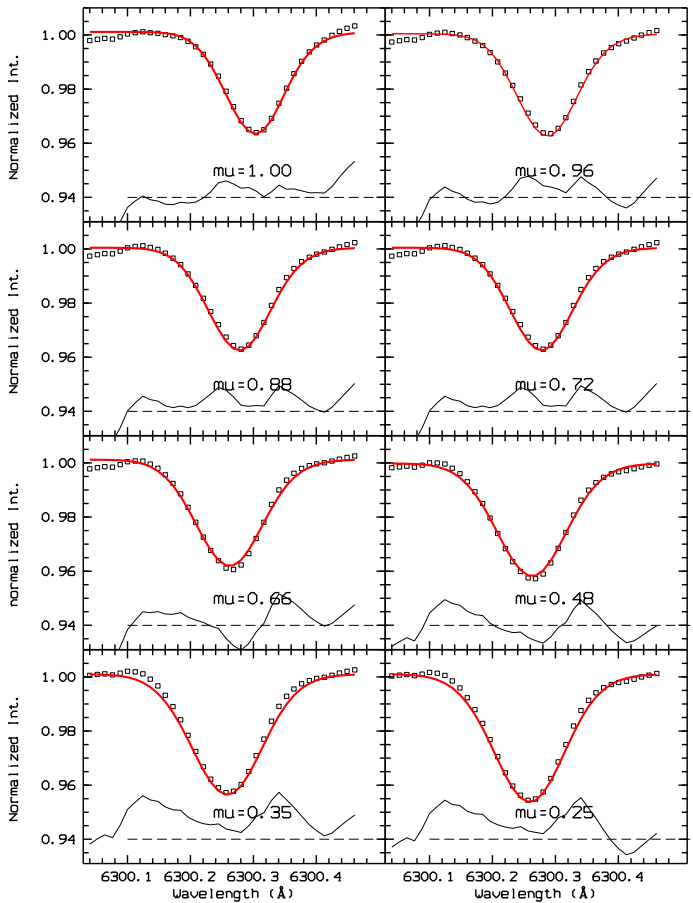}}
\caption{Fitted 3D line profiles at the eight $\mu$-values (red)
 compared to the THEMIS observed spectra (black).
 From the fit we derive A(O)=8.76 and A(Ni)=6.02.
Below each profile the residual ($\times 5$, displaced by +0.94) is shown.}
\label{plotfitthemis}
\end{figure}

As we did in the previous cases, we compared
the observed line profile of the disc-centre spectrum
to the solar atlases of \citet{delbouille} and \citet{neckelobs}.
From this comparison we estimate a scattered light fraction of about 4\%.
The quality of the THEMIS spectra is inferior to that of the other sets 
of observations, so that the continuum is difficult to define.
This could be due to a not entirely successful flat-field correction 
of the data. From the fit shown in Fig.\,\ref{plotfitthemis} we obtain 
A(O)=8.76 and A(Ni)=6.02, but the fit is not as good as for the other 
observations.

\subsubsection{Monte Carlo tests}

We investigated the role of the S/N ratio on the uncertainty
in the abundance determination.
We performed two different Monte Carlo simulations. 
In the first, we injected Poisson noise in the synthetic spectra
with parameters A(O)=8.66 and A(Ni)=6.25.
We considered the five $\mu$-values of the \textit{Hinode} data, 1.00, 0.95, 0.86, 0.70, and 0.38.
The results are presented in Table\,\ref{mcsint}.
The uncertainties in the Table are fit-to-fit scatter
for three cases of noise injection of signal-to-noise of 800, 400, and 300.
We can consider these uncertainties as the statistical error in the abundance determinations 
related to Poisson noise. Owing to the high-quality spectra we work with,
we suggest an uncertainty of 0.02\,dex and 0.04\,dex for the derived abundance
of oxygen and nickel, respectively.

\begin{table}
\caption{1000 Monte Carlo simulations on synthetic spectra.}
\label{mcsint}
\begin{center}
{
\begin{tabular}{rrr}
\hline
\noalign{\smallskip}
 S/N & A(O) & A(Ni)\\
\noalign{\smallskip}
\hline
300 & $8.66\pm 0.04$ & $6.23\pm 0.08$\\
400 & $8.66\pm 0.03$ & $6.24\pm 0.07$\\
800 & $8.67\pm 0.02$ & $6.24\pm 0.04$\\
\noalign{\smallskip}
\hline
\end{tabular}
}
\end{center}
\end{table}

We then took the {\em observed } \textit{Hinode} spectra, slightly degrading 
their quality by injecting noise, but obtaining a plausible
different ``realisation'' of an observation in order to verify
whether the result is consistent with the previous exercise of
the Monte Carlo simulation over synthetic profiles.
This Monte Carlo simulation consists of
1000 events where Poisson noise is injected 
into the five \textit{Hinode} observed spectra.
The results are presented in Table\,\ref{mcobs}.
We see that the scatter within the realisations for oxygen is
twice the value derived from synthetic spectra, which is expected
because  we start from an observed spectrum.
For Ni the fit-to-fit scatter is about a factor of three larger than in the case of the synthetic spectra.
Nickel seems to be more difficult to derive from observed spectra
because the line is weaker, or perhaps  because there are 
uncertainties in the wavelengths of the Ni isotopic components (see \citealt{rosberg93}).
The results of the Monte Carlo tests give an indication that the systematics dominate the error budget.

\begin{table}
\caption{1000 Monte Carlo simulations on the \textit{Hinode} spectra.}
\label{mcobs}
\begin{center}
{
\begin{tabular}{rrr}
\hline
\noalign{\smallskip}
 S/N & A(O) & A(Ni)\\
\noalign{\smallskip}
\hline
300 & $8.71\pm 0.06$ & $6.13\pm 0.14$\\
400 & $8.71\pm 0.06$ & $6.14\pm 0.14$\\
800 & $8.70\pm 0.05$ & $6.15\pm 0.13$\\
\noalign{\smallskip}
\hline
\end{tabular}
}
\end{center}
\end{table}


\section{Discussion}

\subsection{The blending nickel line}

In the previous sections we have discussed the role of the nickel 
blend in terms of the Ni abundance. However,
we believe it is unlikely that a re-analysis of the solar Ni abundance will
result in a large downward revision.  The recent result of
\citet{scott14}, who find $\mathrm{A(Ni)}=6.20 \pm 0.04$, corroborates this
notion. For the meteoritic nickel abundance, \citet{Lodders} recommends
A(Ni)$=6.22\pm 0.03$, thus a nickel abundance as low as around 6.00 (THEMIS)
would be in stark disagreement with the meteoritic nickel abundance.  
Likewise, such a revision cannot be due to errors in the $f$-value of the 
\ion{Ni}{i} line, which is known with an uncertainty of only 
14\% \citep{Johansson}, i.e. 0.06\pun{dex} in logarithmic
abundance.  On the other hand, we note that the Ni line has a rather high
excitation energy (4.266\pun{eV}) and that no calculations exist concerning
the deviations from LTE for Ni. NLTE effects may produce a line weaker than
predicted by our LTE computations. However, \citet{scott09} obtained a
rather homogeneous Ni abundance in LTE from all \ion{Ni}{i} lines used in their
analysis, which would make strong departures from LTE for the blending Ni line
rather surprising.
Computations on a \ion{Fe}{i} line with similar atomic properties provided
by Lyudmila Mashonkina (priv. comm.) suggest that the NLTE correction of 
the \ion{Ni}{i} line cannot be the solution to the problem.

We consider here  a Ni abundance of about 6.1\,dex, which -- taking into
account the uncertainty in the oscillator strength of the Ni line -- is
perfectly consistent with the \citet{scott09} result.  We would  like to emphasise
that a low contribution of nickel to the 630\pun{nm} blend corresponding to
$\mathrm{A(Ni)}\approx 6.1$ is also clearly compatible with previous
findings. Combining the result of \citet{ALA01}, $\log
gf\epsilon_\mathrm{Ni}=3.94$, with the presently accepted value of the
oscillator strength ($\log gf_\mathrm{Ni}=-2.11$) results in a nickel
abundance of 6.05. We note that \citet{ALA01} and also 
\citet{scott09} applied the same 3D model atmosphere in their analyses,
but Scott et al. arrived at a significantly higher nickel abundance from
the unblended Ni lines.  \citet{melasp08} used the [OI] line at 557.34\pun{nm}
to obtain an oxygen abundance of $8.71 \pm 0.07$, which is in reasonable
agreement with our result for the two other forbidden lines within the
uncertainties.  \citet{ayres08} analysed the 630\pun{nm} blend with a single
snapshot from the same solar \cobold\ model as used in this work. He obtains
$\mathrm{A(O)}=8.81 \pm 0.04$ and $\mathrm{A(Ni)}=6.08$ (he states a
$\approx 30\,\%$ smaller Ni abundance than the value given by
\citet{grevesse98} which is $6.23 \pm 0.04$) adopting a rather large
equivalent width for the feature at disc-centre of 0.476\pun{pm}, and applying
corrections to the model to reproduce the measured continuum intensity at the
wavelength of the blend. While his result is certainly at the high end at the
presently discussed oxygen abundances, again, the contribution to the blend by
nickel is small.

\subsection{The role of the centre-to-limb variation}

One should bear in mind that the purpose of this work was to investigate the
line formation in the solar photosphere at different limb-angles. In this
respect the data we used are unique, but not all these spectra have the high
quality of the traditional solar atlases. As we pointed out above, the 
equivalent widths (EWs) of the two WCL and THEMIS 
disc-centre spectra of the 630\pun{nm} line are
smaller than the ones of the largely used disc-centre solar atlases
\citep{delbouille,neckelobs}.  However, we argue that our observations in
their ensemble are surely sufficient to allow a differential study of the
centre-to-limb variation of the forbidden oxygen feature.

As pointed out in Section\,\ref{vttanalysis}, the synthetic profiles are too
broad when compared to high-quality data of extremely high spectral 
resolution. The extra broadening in the synthetic spectra could indicate 
a  velocity field  in the hydrodynamical model that is too strong. This effect is most 
evident in the VTT observation owing to their high resolution, but the effect 
that the FWHM of the observed spectrum becomes increasingly smaller than 
in the synthesis towards the limb is visible in the SST 
(see Fig.\,\ref{plotfitpereira}) and THEMIS data as well (see
Fig.\,\ref{plotfitthemis}.) 

\section{Conclusions}
 Our results can be summarised as follows:
 
\begin{enumerate}
\item
The oxygen abundance we derive from the 630\pun{nm} blend is 
A(O)=$8.73\pm 0.02\pm 0.05$ (with error estimates from the scatter of the 
different observations and the Monte Carlo simulations).
\item
The above oxygen abundance from the 630\pun{nm} feature is inconsistent with a
Ni abundance of 6.25 and the assumption of LTE: we advocate a lower
contribution of the Ni line to the blend corresponding to a nickel
abundance of $6.11\pm 0.04$, or $\log gf\epsilon_\mathrm{Ni}=4.00$.
\item
The synthetic line profile we obtain from the 3D model is  broader than
expected, and this is particularly evident in comparison with the VTT 
observations with the highest spectral resolving power of 700000.  
This effect, which is not seen for other lines, can be due to an overly large velocity field 
in the model atmosphere, although this effect may also indicate errors in the adopted atomic line
wavelength difference between the two components of the [OI]+\ion{Ni}{i} blend. 
\item
The observed spectra at the lowest limb angles ($\mu < 0.2$) are not reproduced
satisfactorily by our computations.  We speculate that this problem is due 
to the solar model that is neither taking into account sphericity effects nor a chromosphere. 
\item
On the other hand, the synthetic spectra reproduce
most of the observed data for $\mu > 0.2$ very well,
even though the agreement between observed and synthetic 
spectra deteriorates somewhat towards the limb.
\item
With the low oxygen abundance we derive in this work from the [OI] line at 630\,nm,
the discrepancy of about 0.1\,dex with the oxygen abundance derived from 
the [OI] 636\pun{nm} line in Paper\,I and discussed in Paper\,II is reduced to 0.06\,dex due to a lower A(Ni), 
but remains unexplained.
\end{enumerate}


\begin{acknowledgements}
This project was funded by FONDATION MERAC.  
We are grateful to Lyudmila Mashonkina for providing
us with results of her NLTE computations of an \ion{Fe}{i} line
that should be analogous to the \ion{Ni}{i} line discussed in the
present paper.  
\textit{Hinode} is a Japanese
 mission developed and launched by ISAS/JAXA, collaborating with NAOJ as a
 domestic partner, NASA and STFC (UK) as international partners. Scientific
 operation of the \textit{Hinode} mission is conducted by the \textit{Hinode}
 science team organised at ISAS/JAXA. This team mainly consists of scientists
 from institutes in the partner countries. Support for the post-launch
 operation is provided by JAXA and NAOJ (Japan), STFC (UK), NASA, ESA, and NSC
 (Norway). The LARS instrument at the VTT was funded by the 'Senatsauschuss 
 Wissenschaft' of the Leibniz-Gemeinschaft, Germany. EC and HGL acknowledge 
 financial support by the Sonderforschungsbereich SFB881 ``The Milky Way 
 System'' (subprojects A4) of the German Research Foundation (DFG).
EC and PB acknowledge support from Programme  National de Physique Stellaire
of the Institut National de Sciences de l'Univers of CNRS.

\end{acknowledgements}

\balance

\bibliographystyle{aa}

\end{document}